\documentclass{IEEEoj}
\usepackage{amsmath,amssymb,amsfonts}
\usepackage{algorithmic}
\usepackage{graphicx,color}
\usepackage{textcomp}
\usepackage{siunitx}
\usepackage{csquotes}
\usepackage{multirow}
\usepackage{orcidlink}
\usepackage{multirow}
\usepackage{booktabs}

\usepackage{cleveref}
\crefname{section}{Section}{Sections}
\crefname{subsection}{Sub-Section}{Sections}
\crefname{figure}{Fig.}{Fig.}
\crefname{table}{Table}{Tables}
\crefname{equation}{Equation}{Equations}

\usepackage{mwe}
\usepackage{amsmath,amsfonts}
\usepackage{siunitx}
\usepackage[
        defernumbers=true,
        sortcites,
        backend=biber,
        bibencoding=utf8,
        natbib=true,
        hyperref=true,
        backref=false,
        urldate=long,
        style=ieee,
        isbn=false,
        dashed=false,
        url=true,
        eprint=false,
        maxnames=3,
        sorting=none
    ]{biblatex}
\usepackage{balance}
\def\BibTeX{{\rm B\kern-.05em{\sc i\kern-.025em b}\kern-.08em
    T\kern-.1667em\lower.7ex\hbox{E}\kern-.125emX}}
\makeatletter
\def\ps@headings{\def\@oddhead{\vbox{\vspace{17pt}\hsize\textwidth\hbox{\rfxfont\rightmark\hfill}\hfill\par
\smallskip\noindent\hbox to \textwidth{\vrule width\textwidth height.3pt depth0pt}}}%
\def\@evenhead{\vbox{\vspace{17pt}\hsize\textwidth\hfill\hbox{\hfill\rhfont\leftmark}\par
\smallskip\noindent\hbox to \textwidth{\vrule width\textwidth height.3pt depth0pt}}}%
\def\@oddfoot{\hfill\rffont\thepage}\def\@evenfoot{\rffont\thepage\hfill}}
\def\ps@plain{\def\@oddhead{\vbox{\vspace{17pt}\hsize\textwidth\hbox{\rhfont\leftmark\hfill}\hfill\par
\smallskip\noindent\hbox to \textwidth{\vrule width\textwidth height.3pt depth0pt}}}%
\def\@evenhead{\vbox{\vspace{17pt}\hsize\textwidth\hfill\hbox{\hfill\rhfont\leftmark}\par
\smallskip\noindent\hbox to \textwidth{\vrule width\textwidth height.3pt depth0pt}}}%
\def\@oddfoot{\hfill\rffont\thepage}\def\@evenfoot{\rffont\thepage\hfill}}
\makeatother
\AtBeginDocument{\definecolor{ojcolor}{cmyk}{1,0.45,0,0.18}%
\definecolor{ojcolor2}{cmyk}{0,0.91,0.81,0.19}%
\pagestyle{headings}\thispagestyle{plain}}
\begin{document}
\receiveddate{XX Month, XXXX}
\reviseddate{XX Month, XXXX}
\accepteddate{XX Month, XXXX}
\publisheddate{XX Month, XXXX}
\currentdate{XX Month, XXXX}
\doiinfo{JXCDC.2023.3238030}

\title{\textcolor{ojcolor2}{Self-Heating and Parasitic Effects in Multi-Tier CFET Design}}

\author{
SUFIA SHAHIN$^{1,2,\orcidlink{0009-0004-0134-980X}}$ (Member, IEEE), 
MAHDI BENKHELIFA$^{1,\orcidlink{0000-0002-8982-2902}}$, 
YOGESH SINGH CHAUHAN$^{2,\orcidlink{0000-0002-3356-8917}}$ (Fellow, IEEE),
HUSSAM AMROUCH$^{1,\orcidlink{0000-0002-5649-3102}}$ (Member, IEEE)}
\affil{Technical University of Munich; TUM School of Computation, Information and Technology, Chair of AI Processor Design, \\Munich Institute of Robotics and Machine Intelligence, Technical University of Munich, 80333 Munich, Germany}
\affil{Department of Electrical Engineering, Indian Institute of Technology (IIT) Kanpur, 208016, India}
\corresp{CORRESPONDING AUTHOR: SUFIA SHAHIN (e-mail: sufia.shahin@tum.de)}

\begin{abstract}
In this article, we study the impact of self-heating effects (SHEs) and middle of line (MOL) and back-end of line (BEOL) induced parasitics on multi-tier CFET design, where multiple nanosheet devices are vertically stacked. We analyze and compare the 4-tier CFET design with the conventional 2-tier CFET, using TCAD models calibrated to experimental measurements. 
Additionally, TCAD simulations are used to model and analyze SHE-induced heat distribution and temperature profiles and to extract the detailed parasitic RC network from 3D models of CMOS inverters designed with full MOL and BEOL interconnects. 
At the device level, the maximum temperature rise ($\Delta$T\textsubscript{MAX}) caused by SHE in nFET and pFET devices of the 2-tier CFET architecture is \SI{62}{\kelvin} and \SI{74}{\kelvin}, respectively. Due to the increased distance from the substrate heat sink, the upper-tier nFET and pFET devices in the 4-tier design show higher $\Delta$T\textsubscript{MAX} of \SI{83.5}{\kelvin} and \SI{98.5}{\kelvin} and more heat trapping in the stacked layers.
Furthermore, in the 4-tier CFET-based CMOS inverters, the BEOL-induced parasitic RCs are, respectively, 10 and 6.5 times higher in the top-tier than in the 2-tier CFET-based inverters. In the bottom tier, the corresponding parasitic RC elements are 6.26 and 2 times higher, respectively, than in the 2-tier inverters.
Finally, compared to the 4-tier design without parasitics, the propagation delay of the top and bottom tier inverters increases by 10\% and 8.2\%, respectively, due to the interconnect parasitic RCs. For the conventional 2-tier inverter, the corresponding degradation of delay with parasitic RCs is 37.25\%.
\end{abstract}


\maketitle

\section{INTRODUCTION}\label{seC:intro}
\IEEEPARstart{T}{HE} Complementary FET (CFET) proposed at the sub-\SI{3}{\nano\meter} technology node has received extensive attention due to its high integration density~\cite{7358115,8993525,cheng2021complementary}. 
The vertical stacking of Gate-All-Around (GAA) nanosheet-based nFET and pFET transistors in the CFET architecture considerably reduces the n-/pFET spacing. This leads to up to 50\% reduction in standard cell area in comparison to the conventional CMOS technologies~\cite{9146737,HRICEE}. Further, the decrease in the n-/pFET spacing leads to reduced miller capacitances, improving the circuit-level performance~\cite{imecFSFET,CFETvsNSFET,HRTC,HRTED}.
Recently, the multi-tier CFET architecture has been proposed for continued scaling beyond the A5 node~\cite{IRDS2024,IEDM4T,11130289}. Multi-tier CFET architecture offers performance gains, feasibility of vertical integration, and improved fabrication density to overcome the limitations of traditional CPP scaling~\cite{IEDM4T,PPA}.
However, scaling down CMOS technologies is faced with adverse short channel effects, pronounced process variations, increased leakage, and exacerbated self-heating effects (SHEs) ~\cite{HRJxCDC,10994809,8370780}. 
Self-heating is a serious reliability concern for transistors, as it leads to a reduction in the ON-state current (I\textsubscript{ON}) and transconductance (g\textsubscript{m}). Furthermore, SHE accelerates defect generation and thus leads to more aging~\cite{11215725}.
This causes significant performance degradation at both the device and circuit levels~\cite{7409678}.
In the GAA transistor architectures, the gate stack composed of low-thermal-conductivity materials covers the channels from all sides~\cite{11215725}. This hinders the dissipation of the heat in these devices, causing excessive temperature rise and localized hotspot formation in the channel, leading to severe current degradation due to increased phonon scattering~\cite{1319147,11215725,9490296}. 
Moreover, the stacked architecture of CFET further degrades heat dissipation efficiency and introduces intra-device thermal crosstalk between the nFET and pFET, thus amplifying the overall self-heating and the associated performance degradation~\cite{9633122,10040990}. 
It is hence crucial to analyze the performance degradation related to SHE at the device and circuit levels for the CFET technology for reliable performance~\cite{9490296,8856276}.
\begin{figure}[!t]
\centerline{
\includegraphics[width=0.25\textwidth, height=0.25\textwidth]{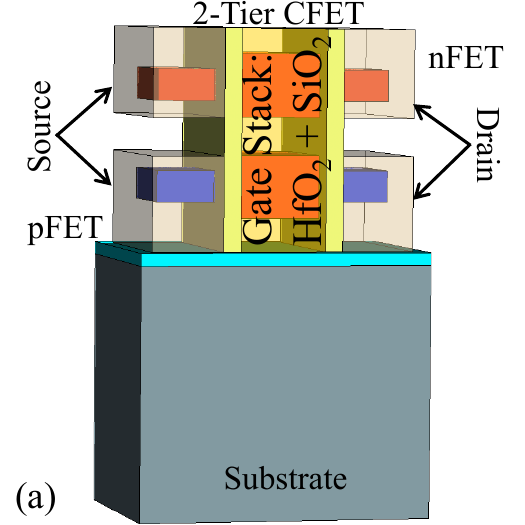}
\includegraphics[width=0.25\textwidth, height=0.25\textwidth]{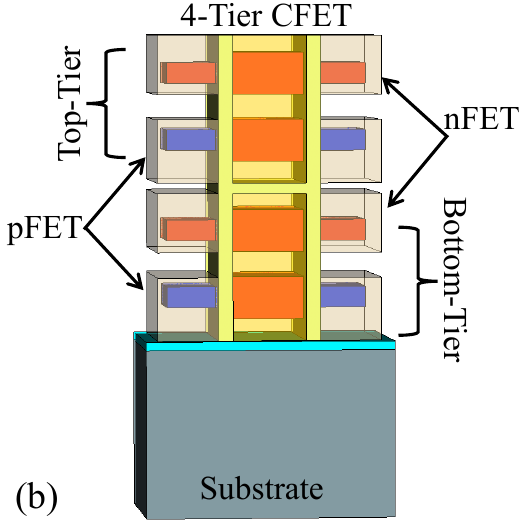}}
\caption{3D TCAD models of the vertically stacked (a) conventional 2-tier CFET and (b) 4-tier CFET devices. Multi-tiering can improve density as an alternative to CPP scaling by means of vertical stacking. The physical dimensions of the TCAD models are listed in \cref{table:parameters}.}
\label{fig:TCAD_models}
\vspace{-2.5mm}
\end{figure}
\begin{figure}[!t]
\centering
\includegraphics[width=1\columnwidth]{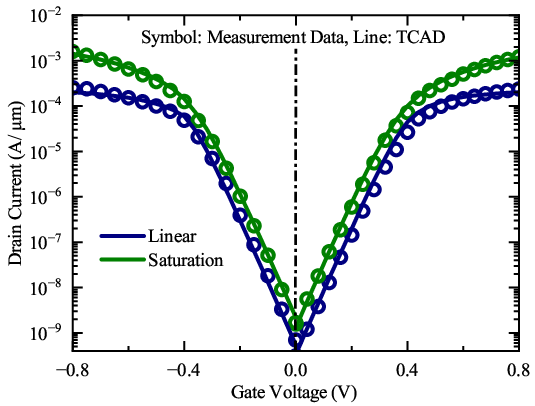}
\caption{The transfer characteristics of our 2-tier CFET model are calibrated in linear (V\textsubscript{DS}=50\,mV) and saturation (V\textsubscript{DS}=0.75\,V) regimes of operation to the experimental measurements~\cite{liao2023complementary}.}
\label{fig:TCAD_Calib}
\vspace{-4mm}
\end{figure}
Furthermore, the circuit performance in advanced technology nodes is also impaired by the parasitic capacitances and resistance of the network of interconnects in the back-end and the middle of line (BEOL and MOL)~\cite{11215725,M3D}. 
Proposed techniques like backside-power-delivery (BPD) separate the power delivery network from the signal routing, alleviating routing congestions in aggressively scaled technology nodes~\cite{cheng2021complementary,11215725,CFETvsNSFET}. However, the study of the impact of SHE and parasitic RCs on multi-tier CFET design remains largely unexplored.

\textbf{Our Key Contributions:} 
In this work, we investigate and compare the impact of SHE as well as MOL- and BEOL-induced parasitic RCs on the conventional and the multi-tier CFET designs using 3D TCAD simulations. First, we develop the 3D TCAD models for the 2-tier CFET and 4-tier CFET devices. 
Next, we calibrate the 2-tier CFET DC characteristics to experimental measurements reported in~\cite{liao2023complementary}. The impact of SHE on device characteristics (I\textsubscript{ON} degradation) and the heatmaps are analyzed through extensive TCAD simulations.
Further, 3D models for the CMOS inverter standard cells are developed in TCAD with full-BEOL interconnects using BPR. 
In \cref{sec:TCAD}, we present the TCAD modeling, calibration, and simulation methodology used to study the impact of SHE. In \cref{sec:Sec2}-\ref{sec:SubSec2a}, the impact of SHE at the device level is analyzed and compared for the 4-tier CFET and the baseline 2-tier CFET. 
Next, in \cref{sec:Sec2}-\ref{sec:SubSec2b}, CMOS inverters with full BEOL interconnects are implemented using the 2-tier  and 4-tier CFET designs, and SHE reliability is compared using TCAD-generated heatmaps and temperature profiles. 
Finally, in \cref{subsec:Sec3}, the MOL- and BEOL-induced parasitic RCs of the 2-tier  and 4-tier CFET-based inverters are compared along with their impact on the performance of these inverters.
\begin{table}
\begin{center}
\centering
\caption{Physical parameters for TCAD modeling and calibration of the 2-tier CFET devices as in~\cite{liao2023complementary}}
\label{table:parameters}
\resizebox{0.8\columnwidth}{!}{
\begin{tabular} {l|c}
\toprule 
\multicolumn{2}{c}{\textbf{TCAD Model Parameters and Dimensions}}\\
\midrule
Gate length                 & \SI{15}{\nano\meter}        \\ 
Sheet width                 & \SI{16}{\nano\meter}         \\ 
Sheet thickness             & \SI{6}{\nano\meter}           \\ 
Effective Oxide thickness   & \SI{0.9}{\nano\meter}          \\ 
Spacer thickness            & \SI{5}{\nano\meter}             \\ 
Channel doping              &   \SI{e15}{\centi\meter}\textsuperscript{-3}     \\ 
Source/Drain doping         & \SI{e20}{\centi\meter}\textsuperscript{-3}       \\ 
Nominal supply voltage                & \SI{0.75}{\volt}      \\ 
\midrule
\multicolumn{2}{c}{\textbf{Physical Parameters for TCAD calibration}}\\
\midrule
Gate Work Function    & \SI{4.55}{\eV}\\
Carrier Mobility      & {600, 470 cm$^{2}$(Vs)$^{-1}$} \\
Saturation Velocity   & {1$\times$10$^{6}$, 6$\times$10$^{5} $ms$^{-1}$}\\
\midrule
\multicolumn{2}{c}{* The paired values correspond to electrons, holes.}
\end{tabular}}
\end{center}
\label{tab1}
\vspace{-5mm}
\end{table}
\section{TCAD Device Modeling and Calibration}\label{sec:TCAD}
The 3D TCAD models for the 2-tier CFET and 4-tier CFET structures are shown in \cref{fig:TCAD_models}. The structural and physical parameters for the physical modeling and DC calibration of the 2-tier CFET are summarized in \cref{table:parameters}. 
We calibrate the device threshold voltage (V\textsubscript{TH}), sub-threshold slope, and Off-state current (I\textsubscript{OFF}) by tuning the gate work function, carrier mobility, and interface-trap concentrations. I\textsubscript{ON} is calibrated using the velocity saturation parameters.
Philips and Lombardi mobility models are adopted to model the carrier mobility and various scattering effects, respectively~\cite{sentaurus_device}. 
Further, Fermi-Dirac statistics and a bandgap narrowing models are incorporated for carrier densities.
For carrier velocity saturation in the presence of high electric fields, we use the high-field saturation model, while the band to-band tunneling (BTBT) model is implemented to characterize carrier tunneling. 
In \cref{fig:TCAD_Calib}, the calibrated transfer characteristics of the 2-tier CFET show decent agreement with the experimental measurements~\cite{liao2023complementary}.

\section{Self-Heating Analysis in Multi-Tier CFET Design}\label{sec:Sec2}
In this section, we discuss the impact of SHE on 2-tier  and 4-tier CFET designs from device to circuit level. At the device level, the maximum temperature increase ($\Delta$T\textsubscript{MAX}) and the percentage degradation in I\textsubscript{ON} are used to evaluate the effect of SHE with each device biased to turn on individually.
For circuit impact of SHE, we examine the heatmaps generated for $\Delta$T\textsubscript{MAX} and heat dissipation in CMOS-inverters implemented with full BEOL-interconnects in TCAD.
\begin{figure}[!t]
\centering
\includegraphics[width=0.50\textwidth, height=0.25\textwidth]{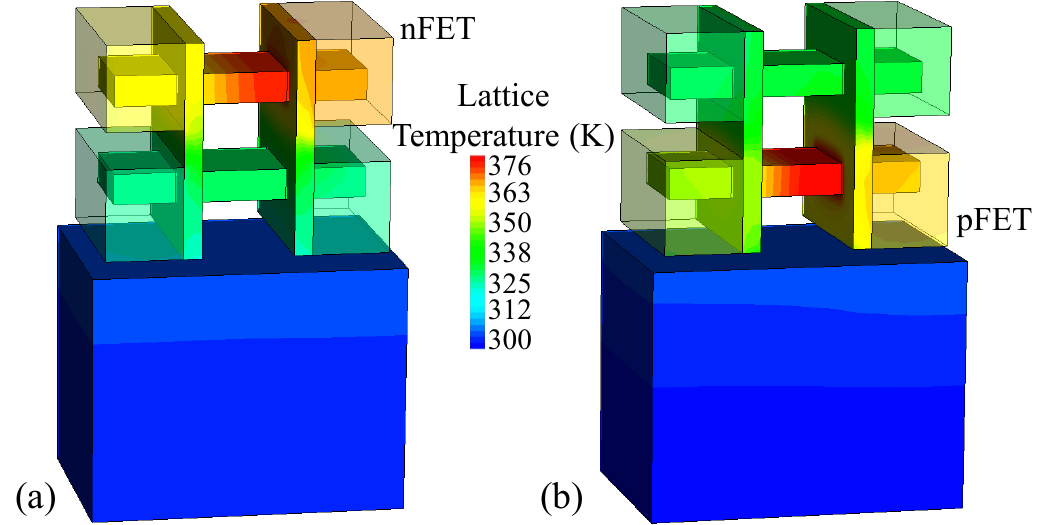}
\caption{Heatmaps generated for (a) nFET and (b) pFET devices, due to SHE, in the 2-tier CFET design extracted using TCAD simulations invoking the thermodynamic transport model and thermal conductivities sourced from~\cite{heat1,heat2}. The devices are biased to turn on individually.}
\label{fig:2TCFET_HM}
\vspace{-4mm}
\end{figure}
\subsection{Impact of SHE at Device Level}\label{sec:SubSec2a}
\begin{figure}[!t]
\centering
\includegraphics[width=0.50\textwidth, height=0.50\textwidth]{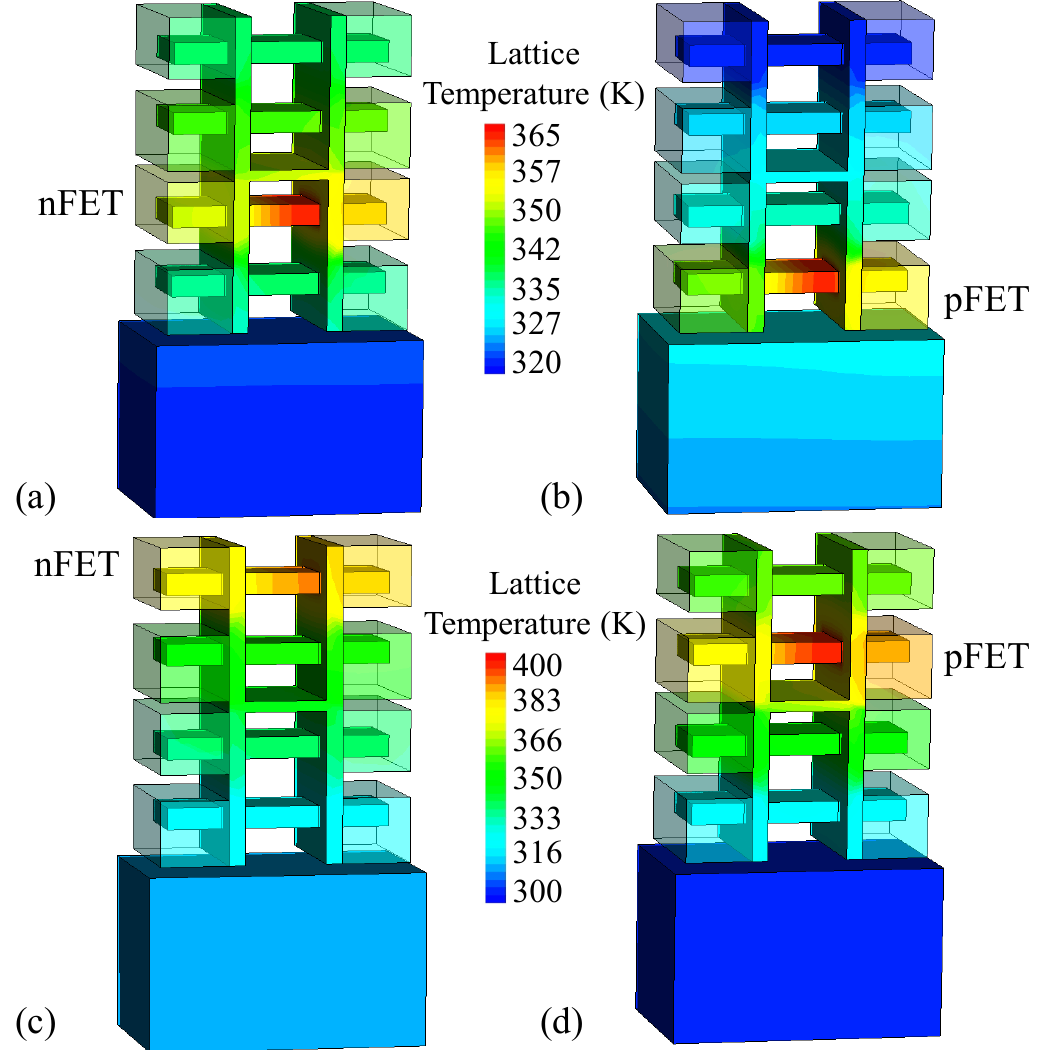}
\caption{SHE induced heatmaps generated for the individually turned on nFET and pFET devices in (a), (b) bottom-tier and (c), (d) top-tier of the 4-tier CFET. Top-tier devices exhibit more heating and rise in temperature than the bottom-tier devices due to a larger distance from the substrate.}
\label{fig:4TCFET_HM}
\vspace{-5mm}
\end{figure}

\begin{figure}[!t]
\centering{
\includegraphics[width=0.24\textwidth, height=0.25\textwidth]{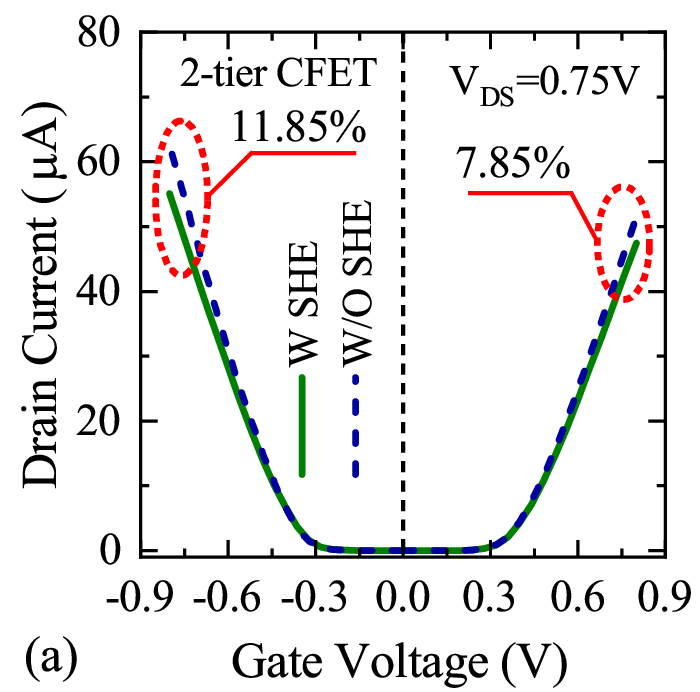}
\includegraphics[width=0.24\textwidth, height=0.25\textwidth]{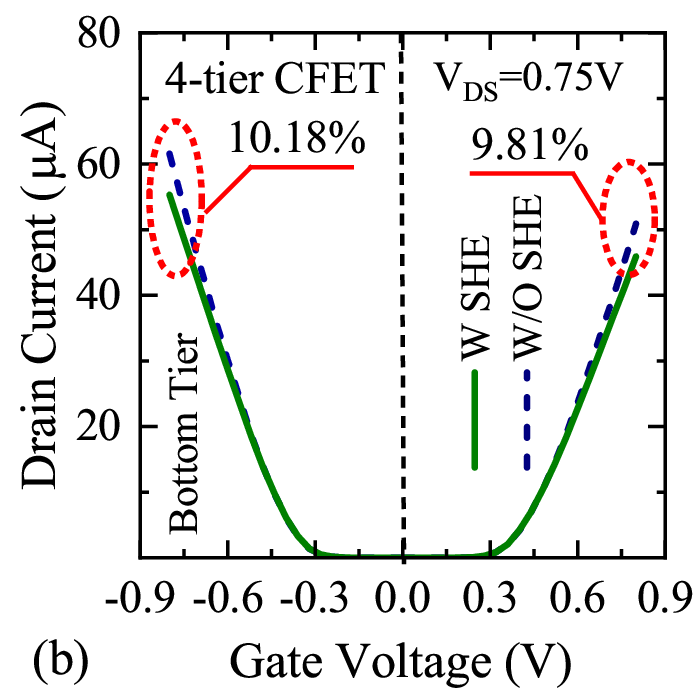}}
\centering{
\includegraphics[width=0.24\textwidth, height=0.25\textwidth]{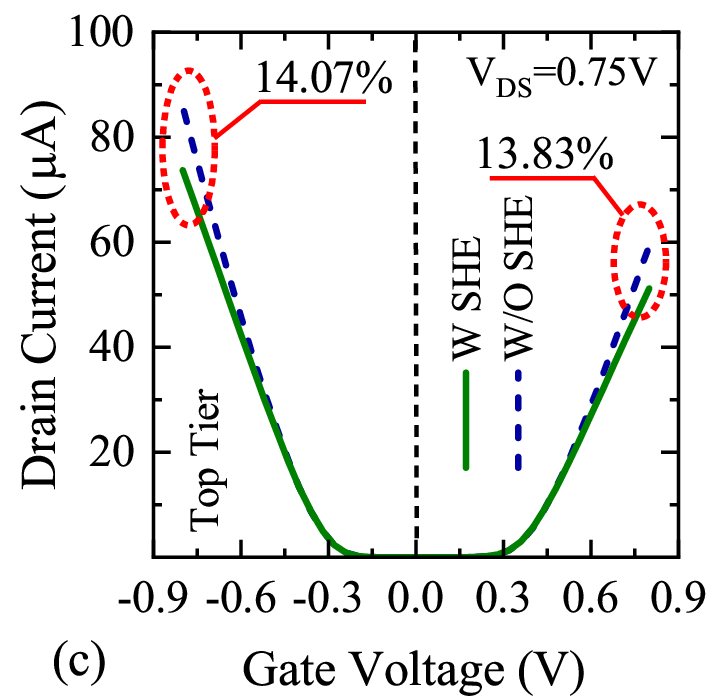}
\includegraphics[width=0.24\textwidth, height=0.25\textwidth]{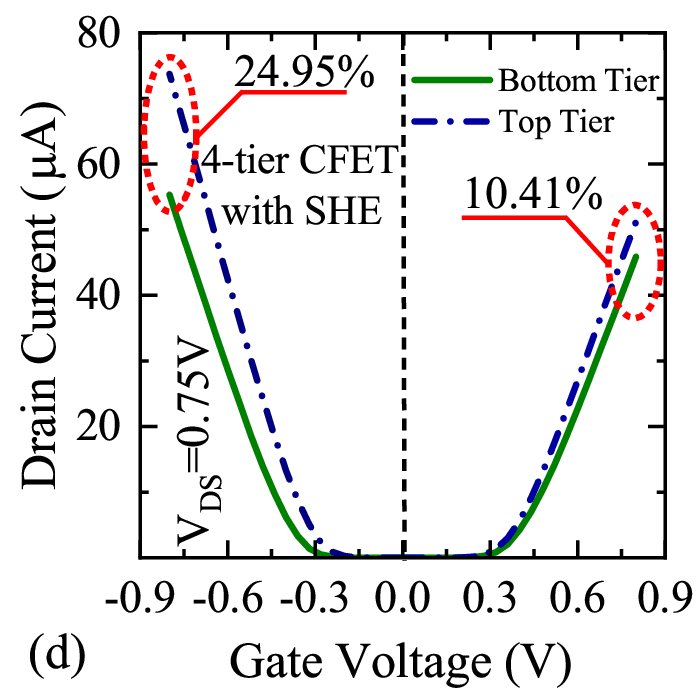}}
\caption{(a) Impact of SHE on 2-tier CFET I\textsubscript{DS}-V\textsubscript{GS} characteristics, with 14\% I\textsubscript{ON} degradation in nFET and 24\% in pFET. Effect of SHE on the DC characteristics of (b) the bottom- and (c) top-tier devices, respectively. (d) Transfer characteristics compared for the bottom- and top-tier devices. Top-tier devices exhibit better drive strength than the bottom-tier devices.}
\label{fig:TCAD_SHE}
\vspace{-4mm}
\end{figure}

To account for the SHE, the thermodynamic transport model is used in conjunction with the temperature-dependent Shockley-Read-Hall (SRH) and Auger models for carrier recombination~\cite{10994809}. Thermal conductivities for materials such as the Si channel, gate, spacer, and oxides sourced from~\cite{11215725}~\cite{heat2}~\cite{heat1}. 
Moreover, proper thermal boundary conditions at the thermal contacts are established to characterize the heat dissipation throughout the structures~\cite{11215725}.
In GAA transistor architectures like CFETs, the channel is surrounded on all sides by a gate stack composed of low-thermal-conductivity insulators such as HfO\textsubscript{2} and SiO\textsubscript{2}. This greatly limits the heat dissipation from the channel region generated due to SHE~\cite{11215725}.
Consequently, in \cref{fig:2TCFET_HM,fig:4TCFET_HM}, the SHE hotspots are localized near the drain region of the devices and concentrated at the center of the nanosheets.
In \cref{fig:TCAD_SHE} (a), we observe that due to SHE, the I\textsubscript{ON} for the 2-tier CFET degrades by 7.85\% for the nFET, while for the pFET, the degradation is 11.85\%. This difference between nFET and pFET in terms of I\textsubscript{ON} degradation in our design is because the I\textsubscript{ON} for pFET is 17.5\% larger than that of the nFET, leading to a higher power density and hotter localized hotspots in pFET~\cite{liao2023complementary,11215725}. 
The corresponding $\Delta$T\textsubscript{MAX} incurred due to SHE by the nFET and pFET devices in the 2-tier CFET design are \SI{62}{\kelvin} and \SI{74}{\kelvin} respectively in \cref{fig:2TCFET_HM}.
This difference in $\Delta$T\textsubscript{MAX} is due to the increased distance between the nFET and the substrate as a result of vertical stacking over the pFET~\cite{11215725}.
Similarly, we observe, from  \cref{fig:TCAD_SHE} (b) and (c), for the bottom and top tier devices in 4-tier CFET-design, the SHE-induced I\textsubscript{ON} degradation for nFET is 9.81\% and 13.83\%, respectively. The corresponding degradations for the pFET devices in the respective tiers are 10.18\% and 14.07\%.
Furthermore, for the 4-tier CFET, I\textsubscript{ON} in the top-tier nFET (pFET) is 10.4\% (25\%) higher than the bottom-tier, \cref{fig:TCAD_SHE} (d), leading to higher temperature rise and more intensified localized hotspots as shown in \cref{fig:4TCFET_HM}(b) and (d).  
In \cref{fig:4TCFET_HM}, $\Delta$T\textsubscript{MAX} in the bottom-tier nFET and pFET devices is \SI{52}{\kelvin} and \SI{61}{\kelvin} respectively, while the corresponding $\Delta$T\textsubscript{MAX} for the top-tier devices is \SI{83.5}{\kelvin} and \SI{98.5}{\kelvin}. Increased distance from the substrate, stronger drive strength, and intra-device thermal coupling are the main causes of higher heating in the top-tier devices.

\subsection{Self-Heating Analysis at Circuit Level}\label{sec:SubSec2b}
\begin{figure}[!t]
\centering
\includegraphics[width=0.5\textwidth, height=0.25\textwidth]{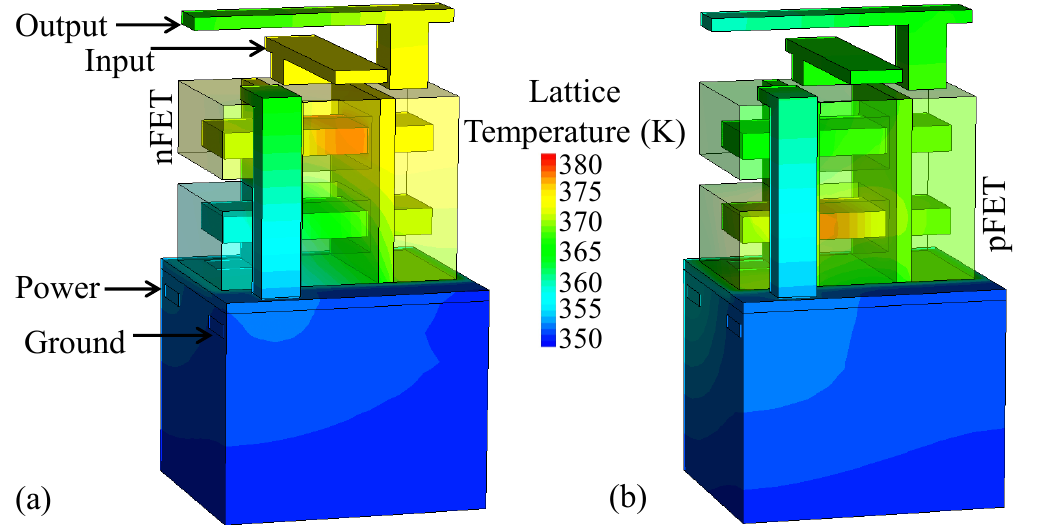}
\caption{Lattice temperature profiles extracted from TCAD simulations for 2-tier CFET based inverter cells with (a) nFET and (b) pFET turned on individually. The $\Delta$T\textsubscript{MAX} for the respective cases is 72\,and 80\,K.}
\label{fig:2TCFET_INV_HM}
\vspace{-5mm}
\end{figure}

\begin{figure}[!t]
\centering
\includegraphics[width=0.50\textwidth, height=0.50\textwidth]{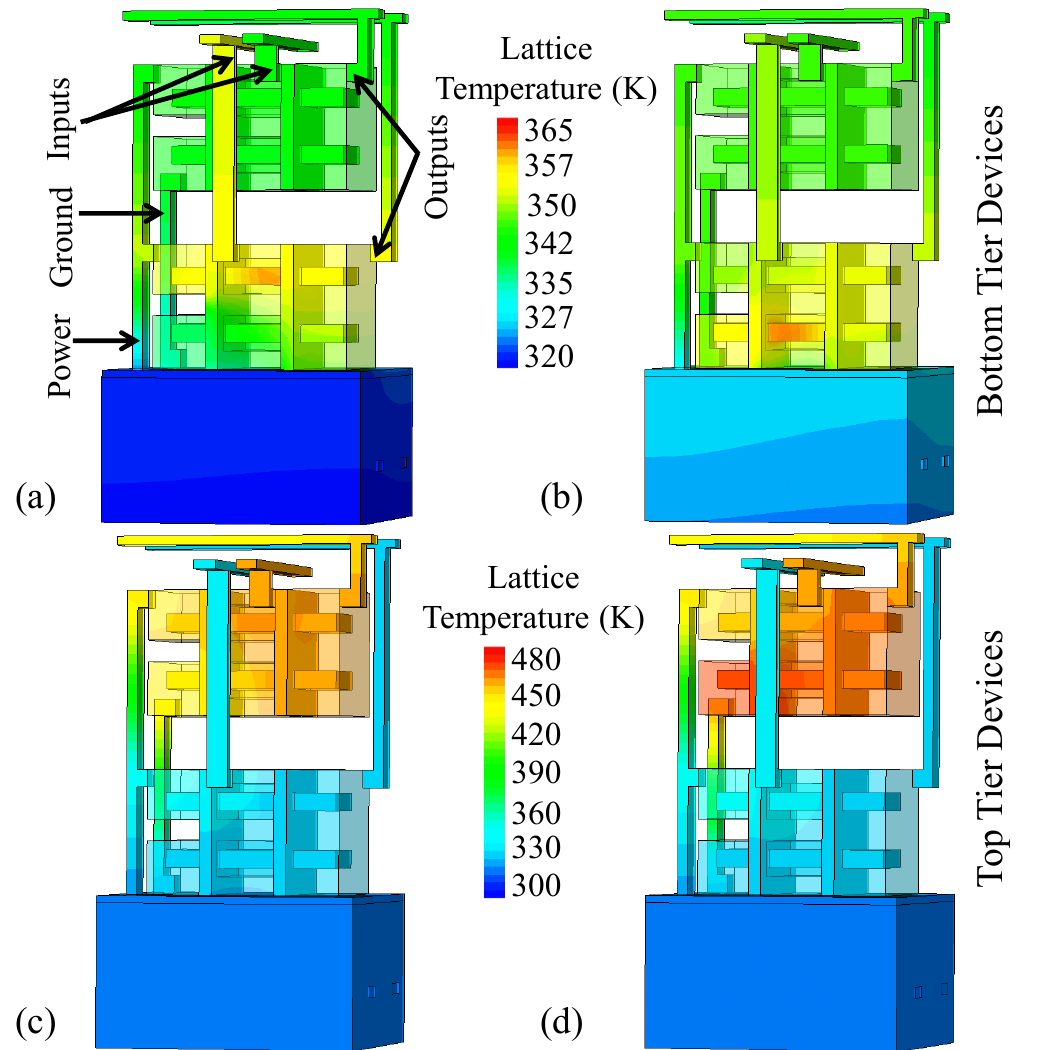}
\caption{The TCAD-generated lattice temperature profiles for the 4-tier CFET inverter design. When (a) bottom-tier nFET, (b) bottom-tier pFET, (c) top-tier nFET, and (d) top-tier pFET devices are turned on independently, the $\Delta$T\textsubscript{MAX} is 58\,K, 61\,K, 162\,K, and 171\,K, respectively.}
\label{fig:4TCFET_INV_HM}
\end{figure}

At the circuit level, the impact of SHE on the 2-tier  and 4-tier CFET-based inverter designs is also assessed using TCAD simulations. 
Detailed 3D TCAD models for the 2-tier  and 4-tier CFET-based CMOS inverters are developed with full BEOL interconnects and buried power rails~\cite{11215725}. To analyze the heat distribution and temperature profiles due to each transistor, the underlying devices are biased to turn on separately through the BEOL contacts. 
The lattice temperature profiles for the 2-tier CFET inverter are shown in ~\cref{fig:2TCFET_INV_HM}. The $\Delta$T\textsubscript{MAX} due to the nFET device in \cref{fig:2TCFET_INV_HM} (a) is \SI{72}{\kelvin}, while for the pFET it is \SI{80}{\kelvin}, in \cref{fig:2TCFET_INV_HM} (b). 
Similarly, the heatmaps of \cref{fig:4TCFET_INV_HM} (a) and (b) show that $\Delta$T\textsubscript{MAX} is \SI{58}{\kelvin} and \SI{61}{\kelvin} for nFET and pFET, respectively, in the bottom-tier. The heat dissipation path from nFET to the substrate is blocked by pFET, while the heat dissipation path from pFET to BEOL is also hindered by nFET to a certain extent.
In \cref{fig:4TCFET_INV_HM} (c) and (d), the respective $\Delta$T\textsubscript{MAX} for the nFET and pFET are \SI{162}{\kelvin} and \SI{171}{\kelvin} in the top-tier. 
Consistent to \cref{sec:Sec2}-\ref{sec:SubSec2a}, the increased separation of the top-tier devices from the substrate, thermal coupling, and higher drive strength leads to higher power densities and $\Delta$T\textsubscript{MAX} than at the bottom-tier.
Further, since most of the heat is generated in the channel near the drain region, it is noted, from \cref{fig:2TCFET_INV_HM,fig:4TCFET_INV_HM}, that the maximum heat dissipation in the BEOL is carried out through the ``Output" and the ``Input" interconnects. This is due to the direct connection of these interconnects to the drain and gate terminals of the underlying transistors. 

\section{Impact of Multi-Tier Design on BEOL Parasitics}\label{subsec:Sec3}

To study the impact of stacking of multiple tiers on the BEOL-induced parasitic RCs and consequent circuit performance degradation, we use the same inverter standard cells as for SHE analysis in \cref{sec:Sec2}-\ref{sec:SubSec2b}. 
The 3D TCAD models of the inverter standard cells are used to extract the parasitic RCs due to the BEOL interconnects using the Synopsys Raphael tool~\cite{b5}. 
The RC elements from the BEOL network for the 2-tier CFET-based inverters are compared individually to the top- and bottom-tier 4-tier CFET design in~\cref{tab1}.
Further, as reported in~\cite{8510618}, the parasitic resistances dominate the performance drop since the deep metal vias from the power rails and top interconnects contribute the largest parasitic resistances. The parasitic capacitances, however, remain in a similar order of magnitude~\cite{8510618}.
We also observe that the parasitic resistances for the 4-tier CFET inverters are in general, larger than those of the 2-tier CFET inverters due to more complex and longer interconnects, \cref{fig:2TCFET_INV_HM,fig:4TCFET_INV_HM}. Moreover, the 2-tier CFET architecture exhibits overall lower parasitic RC as compared to 4-tier CFET. 
Further, within the 4-tier inverter, the top-tier parasitic resistances are the largest since it has the longest interconnect vias, particularly for the ``Ground" and ``Power" rails connection to BPD.
Multi-tier CFET offers improved packing density by stacking multiple layers; however, this is accompanied by an increase in the parasitic capacitances, as reflected in \cref{tab1}. 
In \cref{tab2}, we tabulate the relative increase of different parasitic RC components in the 4-tier design against the respective 2-tier CFET inverters calculated as the ratio of the corresponding parasitic RCs.
We observe from \cref{tab1,tab2} that the average increase in the parasitic resistances and capacitances in the top tier of a 4-tier CFET design is 10 and 6.5 times, respectively, in comparison to a 2-tier CFET. Similarly, the parasitic RC values for the bottom tier are, on average, 6.25 and 2 times higher than those of the 2-tier CFET. 
Furthermore, the impact of these parasitics is a net degradation in the performance of the inverters~\cite{11215725,M3D}.
With no BEOL parasitics considered, the propagation delay (T\textsubscript{P}) for the 2-tier CFET-based inverter is \SI{0.91}{\pico\second}, which degrades to \SI{1.45}{\pico\second} (+37.25\%) in the presence of interconnect-induced parasitics. For the 4-tier CFET based designs, the T\textsubscript{P} for bottom-tier inverter degrades by 8.22\% from \SI{1.35}{\pico\second} to \SI{1.46}{\pico\second} when BEOL-induced parasitic RC are considered. The corresponding T\textsubscript{P} degradation in the top-tier inverter due to interconnect parasitics is by 10\% from \SI{1.43}{\pico\second} to \SI{1.59}{\pico\second}.
Furthermore, with parasitics accounted for, T\textsubscript{P} for the top-tier and bottom-tier inverters in the 4-tier CFET design increases by 9\% and 1.0\%, respectively, compared to the 2-tier CFET design. This is due to the increased parasitic RC values in the 4-tier CFET design as compared to 2-tier CFET.
\begin{table}[!t]
\caption{Prominent BEOL parasitic RC components compared for 2-tier  and 4-tier CFET based CMOS Inverters}
\begin{center}
\resizebox{\columnwidth}{!}{
\begin{tabular}{|c|c|c|c|c|c|}
\hline
S.& Node &Node & 2-Tier & \multicolumn{2}{c|}{4-Tier CFET} \\
\cline{5-6}
No. & 1 & 2 & CFET & Bottom-Tier & Top-Tier\\
\hline
R1 &  Ground & NSource & 21.62 $\Omega$  & 68.18 $\Omega$  & 160.7 $\Omega$ \\
\hline                
R2 & PSource & Power  & 11.37 $\Omega$ & 48.76 $\Omega$  & 120.5 $\Omega$ \\
\hline                
R3 & Input & Gate     & 5.61 $\Omega$  & 78.12 $\Omega$  & 11.72 $\Omega$ \\
\hline                
R4 & Output  & Drain  & 3.89 $\Omega$  & 73.48 $\Omega$  & 18.89 $\Omega$ \\
\hline                
C1 & Power & Input    & 5.13E-24 F     & 1.13E-22 F      & 2.71e-23 F     \\
\hline                
C2 & NSource  & Gate  & 1.17E-17 F     & 1.25E-17 F      & 1.36E-17 F     \\
\hline                
C3 & Gate  & Drain    & 2.66E-17 F     & 3.01E-17 F      & 2.88E-17 F     \\
\hline                
C4 & Output  & Gate   & 2.29E-19 F     & 2.58E-20 F      & 1.66E-20 F     \\
\hline
\end{tabular}}
\vspace{-5.00mm}
\label{tab1}
\end{center}
\end{table}
\vspace{-2.5mm}
\section*{Conclusion}\label{sec:Conclusion}
In this work, we analyze the effect of multi-tier stacking of CFET on SHE reliability, MOL- and BEOL-induced parasitics, and performance from device to circuit level. 
The stacked architecture of the 2-tier CFET exhibits typical hotspots centered in the channel, observed in TCAD-generated heatmaps and lattice temperature profiles. 
Further, the intra-device thermal coupling and larger separation from the substrate lead to more heating in the top tier of the 4-tier CFET devices. The overall temperature rise for pFETs is more than the nFETs throughout the 2-tier and 4-tier CFETs due to higher pFET drive strength. In case of CMOS inverters, the temperature rise due to SHE can reach maximum values of \SI{80}{\kelvin} and \SI{171}{\kelvin} in 2-tier  and 4-tier CFET designs, respectively.
Furthermore, the complex routing and longer BEOL interconnects in the 4-tier CFET-based CMOS inverter standard cells lead to overall increased parasitic RC contributions. This directly translates to a larger degradation in the circuit performance. Consequently, T\textsubscript{P} for 4-tier CFET based inverter is on average 4.75\% higher than the baseline 2-tier CFET inverters.
In conclusion, the vertically stacking CFET devices across multiple tiers offers advantages such as higher density and reduced spacing between active devices. However, the increased SHE and BEOL-induced degradation on device and circuit performance necessitates robust circuit design practices to ensure reliable and optimized operation. 
\begin{table}[!t]
\caption{Ratio of BEOL Parasitic RC Components of 4-tier CFET with Respect to 2-tier CFET}
\begin{center}
\resizebox{\columnwidth}{!}{
\begin{tabular}{|c|c|c|c|c|}
\hline
S. & Node & Node & \multicolumn{2}{c|}{Ratio of Parasitic RC Elements} \\
\cline{4-5}
No. & 1 & 2 & (Bottom-Tier vs 2 Tier) & (Top-tier vs 2 Tier) \\
\hline
R1 & Ground  & NSource & 3.15  & 7.43 \\
\hline
R2 & PSource & Power   & 4.29  & 10.60 \\
\hline
R3 & Input   & Gate    & 13.93 & 2.09 \\
\hline
R4 & Output  & Drain   & 18.89 & 4.86 \\
\hline
C1 & Power   & Input   & 21.87 & 5.28 \\
\hline
C2 & NSource & Gate    & 1.07  & 1.16 \\
\hline
C3 & Gate    & Drain   & 1.13  & 1.08 \\
\hline
C4 & Output  & Gate    & 0.11  & 0.07 \\
\hline
\end{tabular}}
\vspace{-5.00mm}
\label{tab2}
\end{center}
\end{table}

\section*{Acknowledgement}
We acknowledge Swati Deshwal from the Technical University of Munich, Germany, for support during the initial development of SHE TCAD framework.
\balance
\printbibliography
\end{document}